\documentclass[prd,onecolumn,notitlepage,nofootinbib,superscriptaddress,showpacs,showkeys]{revtex4-2}

\usepackage{amsfonts}
\usepackage{graphicx}
\usepackage{dcolumn}
\usepackage{amsmath}
\usepackage{amssymb}
\usepackage[T1]{fontenc}
\usepackage[utf8]{inputenc}
\usepackage{esint}
\usepackage[brazil,english]{babel}
\usepackage[usenames,dvipsnames]{color}
\usepackage{longtable}
\usepackage{ulem}
\usepackage{nicefrac}
\usepackage{bm}
\usepackage{tikz}
\usepackage{braket}
\usepackage{graphicx, nicefrac}
\usepackage{hyperref}
\usepackage{subfloat}

\usetikzlibrary[shapes,arrows]

\usepackage{cancel} 
\usepackage{ulem} 

\newcommand{\p}{\partial}

\begin{document}

\title{Hawking Atmosphere of Anti-de Sitter Black Holes}

\author{A. F. Cardona}
\email{afcardonaji@gmail.com}
\author{C. Molina}
\email{cmolina@usp.br}
\affiliation{Universidade de S\~{a}o Paulo, Escola de Artes, Ci\^{e}ncias e Humanidades, Avenida Arlindo Bettio 1000, CEP 03828-000, S\~{a}o Paulo-SP, Brazil}

\begin{abstract}

This work investigates the semiclassical evolution of the Hawking atmosphere surrounding evaporating, spherically symmetric anti-de Sitter (adS) black holes. We model the evaporation process within a dynamical framework, treating the emission of Hawking radiation as a quantum tunneling process through the black-hole horizon. Using the Parikh--Wilczek tunneling method, we incorporate backreaction effects, with the emission probability being linked to the resulting change in the Bekenstein--Hawking entropy of the black hole. This probability is then used to compute the time-dependent luminosity of the system, revealing significant deviations from ideal blackbody behavior, particularly for small adS black holes. For these objects, the luminosity does not increase with temperature due to strong mass variations during evaporation. To complement this microscopic approach, we compute the renormalized energy--momentum tensor for a quantum field propagating in the Vaidya-adS geometry modelling the evaporation process. Together, these approaches clarify the interplay between geometry, quantum fields, and thermodynamics in shaping the Hawking atmosphere and the evaporation dynamics of black holes in asymptotically adS spacetimes.

\end{abstract}


\maketitle

\section{Introduction}

In recent decades, anti-de Sitter (adS) geometries have emerged as an important topic in theoretical physics. Much of this interest is driven by the adS/CFT correspondence, which provides a holographic bridge between bulk gravity and boundary conformal field theories \cite{Gubser:1998bc,Maldacena:1997re,Witten:1998qj}.
However, adS spaces are interesting in their own right. Unlike their asymptotically flat counterparts, adS black holes exhibit a distinctive evaporation process, with reflecting boundary conditions at conformal infinity leading to a dynamical equilibrium between the emission and absorption of Hawking radiation. This boundary interaction allows for a sharp distinction between the thermodynamicly unstable ``small'' adS black-hole regime and the stable ``large'' adS regime \cite{hawking1983thermodynamics}. 
The unusual thermal properties of adS black holes lead to more complex thermodynamic descriptions of these objects. This has prompted various initiatives to better characterize their thermodynamic properties \cite{hawking1999rotation,gpp,kastor2009enthalpy,dolan2011pressure,Page2018,bfm2,fbfm,Campos:2024fpf,xiao2024extended,Campos:2025nnk}.

Despite extensive study, several important questions about the evaporation process of black holes remain unanswered.  While a complete description requires a fully self-consistent solution to the backreaction problem, a semiclassical treatment provides significant insight. In this approach, the semiclassical properties of black holes are considered in relation to the quantum fields surrounding them, referred to as the ``Hawking atmosphere'' \cite{hawking1975particle,t1998self,Wald:1999vt,hemming2007thermodynamics}.

In this work, we investigate the Hawking atmosphere around adS black holes as an inherently dynamic structure. Although much of our understanding of black-hole radiation stems from the analysis of stationary geometries, the evaporation process is non-static. As particles are radiated away, the black hole’s mass decreases, creating a backreaction that continuously alters the surrounding geometry \cite{massar1995semiclassical}. To capture the physics of an evaporating black hole, one approach is to move beyond static approximations and adopt a framework that accounts for the time-dependence of the black-hole mass and horizon. To model this evolution, we use the Vaidya-adS metric \cite{vaidya1951gravitational, wang1999generalized, xiang2000entropy, hiscock1981models, hiscock1981modelsII, campos2025dynamical, Campos:2024wcq}, providing a framework to track the evolving horizon and study the transition between different thermodynamic states in a  time-dependent manner.

To study the time-dependent evolution of the Vaidya-adS geometry, we employ the Parikh--Wilczek tunneling method \cite{parikh2000hawking,jun2006tunnelling}, which describes evaporation as the tunneling of quantum particles through the black-hole horizon. This approach, directly linking emission to changes in the Bekenstein--Hawking entropy, being particularly well-suited for non-static backgrounds as it bypasses the need for a global definition of a vacuum state, focusing instead on the local geometry of the horizon. 
Complementing this analysis, we consider the renormalized energy--momentum tensor within an optical geometry approximation~\cite{ref-birrell}, providing insights into the behavior of the quantized energy density during the decay. 

\newpage

The remainder of this paper is organized as follows. 
In Section~\ref{section2}, we introduce the key elements of the model for the Hawking atmosphere of a spherically symmetric, asymptotically anti-de Sitter black hole, reviewing the Vaidya-adS spacetime and its thermodynamic properties.
Section~\ref{section3} details the dynamical mechanism of the Hawking atmosphere by applying the tunneling method to this non-static background. 
Section~\ref{section4} complements the analysis with results for the renormalized energy--momentum tensor for a quantum field propagating in the Vaidya-adS geometry modelling the evaporation process. 
Concluding remarks are presented in Section \ref{sec:conclusion}.
In this work, we adopt the metric signature $(-,+,+,+)$ and work in natural units in which $c = G = \hbar = 1$.

\section{Modeling the Hawking Atmosphere}
\label{section2}

\subsection{Dynamical Anti-de Sitter Black Holes}

The simplest asymptotically anti-de Sitter black hole solution is the Schwarzschild anti-de Sitter (SadS) spacetime. In four dimensions, the SadS line element in standard coordinates $(t, r, \theta, \phi)$ is: 
\vspace{-11pt}
\begin{equation}
\label{eq:spherical-static}
{d}s^{2} = - \left( 1 - \frac{2M_{0}}{r} - \frac{\Lambda r^2}{3} \right)
{d}t^{2} 
+ 
\left( 1 - \frac{2M_{0}}{r} - \frac{\Lambda r^2}{3} \right)^{-1} dr^2 + r^{2}\left( {d} \theta^2 + \sin^2 \theta \, {d}\phi^2 \right) \,.
\end{equation}
In Equation~\eqref{eq:spherical-static}, $M_{0}$ is the black hole's mass. 
The metric~\eqref{eq:spherical-static} is a spherically symmetric solution of Einstein's equations in vacuum with a negative cosmological constant $\Lambda$. By the usual extensions of Birkhoff's theorem, SadS spacetime is static. 

The thermodynamic properties of the SadS black hole differ considerably from those of its asymptotically flat counterpart \cite{hawking1983thermodynamics,kastor2009enthalpy,dolan2011pressure,Page2018,bfm2,fbfm}.
A critical feature of asymptotically anti-de Sitter thermodynamics is the existence of a minimum black-hole temperature. This minimum defines a Hawking--Page transition: small black holes exhibit negative specific heat and are globally unstable, whereas large adS black holes possess positive specific heat and are thermodynamicly stable.

Despite advances in thermodynamic investigations of SadS spacetime, the thermal evaporation process of a black hole is fundamentally dynamic. To describe the non-equilibrium Hawking atmosphere of an evaporating, spherically symmetric, asymptotically anti-de Sitter black hole, we use the Vaidya-adS metric \cite{vaidya1951gravitational, wang1999generalized, hobson}. This metric is an exact solution to Einstein's field equations describing an asymptotically anti-de Sitter spacetime surrounding a spherically symmetric body emitting or absorbing null radiation. Its maximal extension can be used to model an evaporating black hole. Vaidya-adS spacetime is not the only dynamical extension of adS geometry, but it is a simple model that captures important aspects of the Hawking atmosphere in spherically symmetric black holes. For instance, this model accounts for the fact that evaporation is a dynamic process dominated by massless particles.

The Vaidya-adS metric is typically expressed in Eddington--Finkelstein coordinates. The outgoing version is based on the coordinates $(u,r,\theta,\phi)$, with $u$ being the retarded (outgoing) null coordinate. In this coordinate system, the Vaidya-adS line element is given by
\begin{equation}\label{eq:metric_Vaidya-adS}
{d}s^{2} = -f(u,r)du^{2} - 2dudr + r^{2}\left( {d} \theta^2 + \sin^2 \theta \, {d}\phi^2 \right) \,,
\end{equation}
where
\begin{equation}
f(u,r) = \Bigg[ 1 - \frac{2M(u)}{r} + \frac{r^2}{L^2} \Bigg]  \,.  
\label{eq:f-vaidya-ads}
\end{equation}
In Equation~\eqref{eq:f-vaidya-ads}, $L$ is the adS radius, defined in terms of $\Lambda$ as $L \equiv \sqrt{-3/\Lambda} \, $.
The outgoing version of the Vaidya-adS metric in Equation~\eqref{eq:metric_Vaidya-adS} can be interpreted as the exterior field of a massive object from which mass is radiating away in the form of pure radiation. Therefore, it is particularly well-suited for describing a black hole emitting particles through evaporation \cite{hiscock1981models, hiscock1981modelsII, campos2025dynamical}.%
\footnote{The ingoing Vaidya-adS sector is usually characterized by a metric using the coordinate system $(v, r, \theta, \phi)$, where $v$ is the advanced null coordinate. In this case, the sign of the second term on the right-hand side of Equation~\eqref{eq:metric_Vaidya-adS} is positive. However, the ingoing Vaidya-adS metric describes a spherically symmetric object absorbing null radiation and is therefore not of primary interest in this work.}

The time-dependency of the metric is encoded in the mass function $M(u)$ of the retarded coordinate $u$. The energy--momentum tensor $T_{\mu\nu}$ supporting this geometry describes a massless null fluid propagating along the $u$ direction:
\begin{equation}\label{eq:energy--momentum_Vaidya}
T_{\mu\nu} = \frac{\dot{M}(u)}{4\pi r^2} l_{\mu}l_{\nu} \,,
\end{equation}
where $\dot{M} \equiv dM/du$ and $l_{\mu} \equiv -\p_{\mu} u$. For an evaporating black hole, $\dot{M} < 0$, implying an outgoing flux of Hawking radiation that effectively reduces the black hole's mass.

Considering dynamical black holes in non-asymptotically flat backgrounds, it is important to establish a proper description of the boundary of the black hole and its associated properties. For this purpose, we follow the formalism presented by Hayward's generalized thermodynamics \cite{Hayward:1994,Hayward:1996,Hayward:1998sh}, in which the black-hole boundary is defined as a future outer trapping horizon, defined as the surface where the expansion of future-directed null geodesics vanishes. For the Vaidya-adS black hole, this occurs at $f(u,r_{H}) = 0$. The solution for the trapping horizon radius, $r_{H}(u)$, is found to be
\begin{equation}\label{eq:trap-horizon-VaidyaadS}
r_{H}(u) = \frac{2L}{\sqrt{3}}\sinh \bigg\{ \frac{1}{3} \text{arcsinh} \bigg[\frac{3\sqrt{3}M(u)}{L} \bigg] \bigg\}  \,.
\end{equation}

Within the framework of generalized thermodynamics, the thermodynamic quantities of the black hole are defined in terms of the trapping horizon. For the Vaidya-adS black hole, the dynamical surface gravity is 
\begin{equation}
\kappa_g(u) 
\equiv 
\frac{1}{2} \p_r f(u,r_{H}) 
= \frac{1}{2} \p_r \left. {\bigg[} 1 - \frac{2M(u)}{r} + \frac{r^2}{L^2} {\bigg]} \right|_{r_{H}}\,,
\end{equation}
from which, in turn, the time-dependent temperature can be obtained
\begin{equation}\label{eq:temperature-adS}
T(u) 
= \frac{\kappa_g(u)}{2\pi} 
= \frac{3r_{H}(u)^2 + L^2}{4\pi r_{H}(u) L^2} \,.
\end{equation}
As long as the mass loss is quasi-adiabatic ($|\dot{M}| \ll r_{H}^{-1}$), this temperature remains a well-defined thermodynamic quantity throughout the evaporation process.

\subsection{Tunneling Mechanism for Evaporation}
\label{sec:Tunneling}

In a semiclassical treatment of black-hole evaporation, the Hawking atmosphere can be modeled with a quantum field propagating in a classical background. The dynamics of matter fields are introduced by a relativistic equation of motion that depends on the curved spacetime background. For our analysis we will focus on the case of a massless scalar field, which provides sufficient simplicity compared to higher spin fields, while at the same time capturing the essential qualitative features that remain representative of more complex physical systems. A massless scalar field $\Phi$ can be described by the Klein--Gordon equation,
\begin{equation}\label{eq:covariant_KG}
\square \Phi = \frac{1}{\sqrt{-g}}\p_{\mu} \left(\sqrt{-g}g^{\mu\nu}\p_{\nu}\Phi \right) = 0 \,.
\end{equation}

The weak-field approximation, considering scalar fields on fixed, curved backgrounds, assumes that second-order corrections to the classical energy--momentum tensor are negligible. This approximation is one of the foundations of black-hole thermodynamics \cite{ref-birrell,hawking1975particle, parikh2000hawking}. 
Investigations of scalar fields within the context of anti-de Sitter backgrounds have provided valuable insights into holographic dualities and the thermodynamics of asymptotically adS spacetimes \cite{Witten:1998qj,Avis:1977yn,Elias:2018yct}.

The classical and semiclassical properties of a scalar field on an adS background are highly nontrivial and differ significantly from those in asymptotically flat geometries. Even in the relatively simpler case of a Schwarzschild–anti-de Sitter black hole, whose metric is given in Equation~\eqref{eq:spherical-static}, the dynamics are governed by an effective potential of the form \cite{horowitz,wang1}
\begin{equation}\label{eq:potential_scalar}
V_{\text{eff}}(r) = \left( 1 - \frac{2M_{0}}{r} - \frac{\Lambda r^2}{3} \right) 
\left[ \frac{\ell(\ell+1)}{r^2} + \frac{2M_{0}}{r^3}
+
\frac{2}{L^{2}}
\right] \,.
\end{equation}
In Equation~\eqref{eq:potential_scalar}, $\ell$ is the multipole number associated with the angular momentum of the field's mode.
This effective potential is positive definite, vanishes at the black-hole horizon, and diverges at the spatial infinity due to the confining nature of adS. Depending on the spacetime parameters, the potential may exhibit a local maximum, which plays a crucial role in determining the black hole's classical and semiclassical characteristics \cite{horowitz,wang1}.
In the dynamical regime, that is, when the spacetime is equipped with the Vaidya-adS metric~\eqref{eq:metric_Vaidya-adS}, the dynamics are no longer characterized by a static potential. An evolving effective potential that generalizes Equation~\eqref{eq:potential_scalar} could be considered, but in this case a full numerical treatment would be unavoidable \cite{abdalla}.

Instead, this work proposes an approach that considers a tunneling framework for characterizing the semiclassical properties of black holes. Specifically, we use the Parikh--Wilczek method to study black-hole evaporation. This technique characterizes Hawking radiation as a semiclassical tunneling effect of particles through the black-hole horizon \cite{parikh2000hawking, jun2006tunnelling}. Specifically, we consider the radial propagation of massless scalar particles through the trapping horizon. It should be noted that the Vaidya-adS is already a first description of the Hawking atmosphere. Nevertheless, we improve upon this description by considering that semiclassical tunneling provides a backreaction contribution to the process. In this sense, the tunneling calculation can be viewed as an additional correction to the evaporation description using the purely classical Vaidya-adS metric.

In the dynamical regime, the Klein--Gordon equation is treated within the eikonal approximation. This approximation retains only high-frequency modes, allowing wave fronts to be effectively described as rays that follow geodesics. Furthermore, the dominant contribution to the tunneling process arises from radial trajectories \cite{parikh2000hawking, visser}.
In this approach, the solutions of Equation~\eqref{eq:covariant_KG} are  expanded in the following form:
\begin{equation}\label{eq:s-waves}
\Phi = \Phi_{0} e^{i S} \,,
\end{equation}
where $S \equiv S(u,r)$ represents the classical action of a massless particle and $\Phi_{0} \equiv \Phi_{0} (u,r)$ is a slowly varying amplitude. For this solution, the field Equation~\eqref{eq:covariant_KG} reduces to the Hamilton--Jacobi equation
\begin{equation}\label{eq:Hamilton--Jacobi}
g^{ab} (\p_{a} S)\, (\p_{b} S ) = 0 \,.
\end{equation}
In Equation~\eqref{eq:Hamilton--Jacobi}, we denote the metric in the normal space to the spheres $S^2$ in Equation~\eqref{eq:metric_Vaidya-adS} by $g^{ab}$ (with Latin indices). Explicitly:
\begin{equation}
g_{ab}
= \begin{pmatrix} 
-f & -2 \\
-2 & 0 
\end{pmatrix}
\Longrightarrow
g^{ab}
= \frac{1}{2}
\begin{pmatrix} 
0 & -1 \\
-1 & f/2
\end{pmatrix}
\, ,
\end{equation} 
with the function $f\equiv f(u,r)$ presented in Equation~\eqref{eq:f-vaidya-ads}.

The action $S$ can be expressed as an integral along a null trajectory $\gamma$ representing the path of massless particles,
\begin{equation}\label{eq:Solution_Hamilton--Jacobi}
S = \int_{\gamma} \p_{a} S dx^{a} \,.
\end{equation}
The key point of the tunneling method proposed by Parikh and Wilczek is that solutions to the Hamilton--Jacobi equation that are allowed classically will yield a real $S$. In the semi-classical extension of the theory, however, $S$ can take on complex values. The emission rate for a particle tunneling through the trapping horizon is then given by
\begin{equation}
\Gamma \sim \Phi^{*} \Phi
\label{eq:tunneling}
\end{equation}
Considering Equation~\eqref{eq:tunneling} and the eikonal approximation, the emission rate scales as
\begin{equation}\label{eq:tunneling_probability}
\Gamma \sim e^{- 2 \, \textrm{Im} S} \,.
\end{equation}

Following the Parikh--Wilczek formalism, the imaginary part of the action of a single particle of energy $\omega$ and momentum $p_r$ that is tunneling from $r_{in}$ (just inside the horizon) to $r_{out}$ (just outside the horizon) is:
\begin{equation}
\text{Im} S = \text{Im} \int_{r_{in}}^{r_{out}} p_r dr 
= \text{Im} \int_{r_{in}}^{r_{out}} \int_{0}^{p_r} dp'_{r}dr
= \text{Im} \int_{r_{in}}^{r_{out}} \int_{0}^{H} \frac{dr \, dH'}{\dot{r}}  \,,
\end{equation}
where the Hamiltonian $H$ is identified with the mass of the black hole, and $\dot{r} = dH/dp_r$ is Hamilton's equation. 

Next, we focus on the Vaidya-adS case.
For radial null geodesics, the equation of motion is given by
\begin{equation}
\dot{r} \equiv \frac{dr}{du} = \frac{1}{2} f(u, r) \,.
\end{equation}    

To account for the dynamical effect of evaporation, we treat the tunneled particle as a shell of energy $\omega$ that modifies the horizon location. The transition occurs between an initial black-hole mass $M$ and a final black-hole mass $M-\omega$. For the Vaidya-adS black hole,
\begin{equation}\label{eq:ImSomega}
\text{Im} S = -\text{Im} \int_{0}^{\omega} d\omega' \int_{r_{in}}^{r_{out}} \frac{2}{1 - \frac{2(M-\omega')}{r} + \frac{r^2}{L^2}} dr \,.
\end{equation}
The integral has a pole at the horizon $r_{H}$. By deforming the contour around the pole (using the Feynman prescription), the residue yields:
\begin{equation}
\text{Im} S = -\frac{1}{2} \Delta S \,, \quad S = \pi r_{H}^2 \,,
\end{equation}
where $\Delta S$ is the change in the Bekenstein--Hawking entropy $S$ of the black hole after the emission of the particle with energy $\omega$. This results in an emission rate given by
\begin{equation}\label{eq:Entropy-distribution}
    \Gamma \sim e^{\Delta S} = e^{S(M-\omega) - S(M)} \,.
\end{equation}

For an evaporating black hole, $\Delta S$ is negative.
This result demonstrates that the tunneling spectrum is not strictly thermal. Rather, it follows a distribution dictated by the change in the available phase space of the black hole \cite{hemming2007thermodynamics}. For small $\omega$, a Taylor expansion of Equation~\eqref{eq:Entropy-distribution} recovers the Boltzmann factor:
\begin{equation}
\Gamma(\omega) \approx \exp\left(-\frac{\omega}{T} + \frac{1}{2}\frac{\p^2 S}{\p M^2}\omega^2\right) \, .   
\label{eq:boltzmann-factor}
\end{equation}
Result~\eqref{eq:boltzmann-factor} provides a consistent bridge between the tunneling picture and the thermodynamic evolution described by the time-dependent temperature.

\section{Mass Evaporation Model}
\label{section3}
In asymptotically anti-de Sitter spacetimes, the causal structure requires specific boundary conditions at infinity. Because adS  geometries are not globally hyperbolic, the information at the conformal boundary must be specified to ensure a well-defined evolution. A crucial consequence of this geometry is that massive particles are trapped by the adS gravitational potential and eventually fall back toward the black hole. Consequently, only massless degrees of freedom contribute to the long-term evaporation process \cite{Page2018}. In this sense, the representation of the evaporation process via tunneling of massless scalar particles is justified.

In order to explore aspects of the time-dependent evolution of the evaporation process of adS black holes, we introduce a model for the time-dependent mass $M(u)$, defined in terms of a hyperbolic tangent sigmoid function:
\begin{equation}\label{eq:mass_sigmoid}
M(u) =  M_{0} + \frac{M_{f} - M_{0}}{2}\left[ 1 + \tanh\left( \frac{u - u_{0}}{b} \right) \right] \,.
\end{equation}
In Equation~\eqref{eq:mass_sigmoid} $M_{0}$ and $M_{f}$ represent the initial and final mass states, respectively. The parameter $b$ dictates the evaporation rate: smaller values of $b$ correspond to a faster transition between mass states. This choice of $M(u)$ provides a $C^\infty$-continuous evolution that avoids the numerical and physical pathologies associated with discontinuous (or non-smooth) models.
We explored some more complex mass function profiles using a fully numerical approach and found that the results were qualitatively similar to those obtained with Equation~\eqref{eq:mass_sigmoid}. An advantage of the the present model is that it allows for a semi-analytical treatment.

Starting from mass evaporation model~\eqref{eq:mass_sigmoid}, we can numerically evaluate the different thermodynamic properties associated with the trapping horizon solution Equation~\eqref{eq:trap-horizon-VaidyaadS}. The numerical results will depend considerably on the relation between the initial mass $M_{0}$ and the adS radius $L$.

\begin{figure}[h]
\includegraphics[width=9.0 cm]{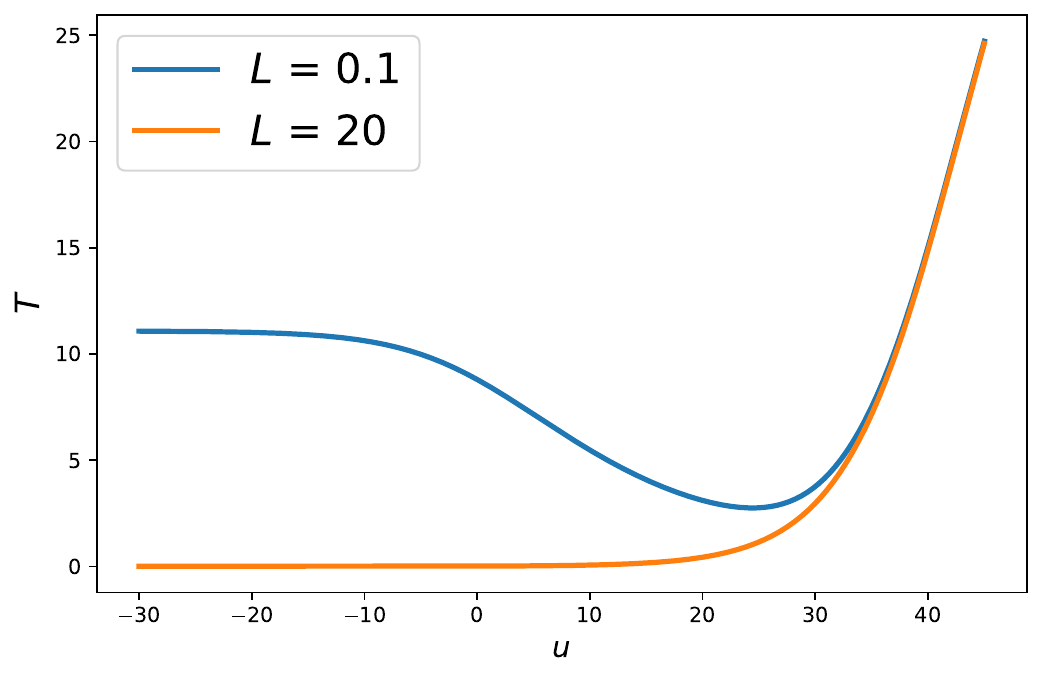}
\caption{Typical temperature profiles for small ($L=20$) and large ($L=0.1$) adS black holes. In this graph, $M_0 = 5$. }
\label{fig1}
\end{figure}

The temperature profile [$T(u)$] in terms of the parameter $u$ reveals the core physics of the adS evaporation cycle. Depending on the adS scale, we observe two distinct behaviors.
In the small black hole regime ($M_0 < L$), the temperature starts at a low value and remains nearly constant during the early stages. Eventually, the temperature rapidly increases, matching the divergent behavior of Schwarzschild-like evaporation. In the large black-hole regime ($M_0 \gg L$), the temperature exhibits an initial decrease, reaching a local minimum before the final runaway heating occurs. This minimum corresponds to the transition point where the black-hole's heat capacity changes sign, moving from the stable ``large'' phase to the unstable ``small'' phase.
Figure~\ref{fig1} illustrates this behavior for the temperature profile. In this graph, the yellow line represents a small adS black hole, beginning with a low temperature that continuously increases as the mass diminishes. The blue line represents a large adS black hole, where the temperature reaches a minimum value before approaching the result of a small black hole.

A key parameter to analyze during the evaporation process of a black hole is the luminosity. The standard Stefan--Boltzmann law assumes the black hole acts as an infinite heat reservoir, staying at a constant temperature $T$ while emitting. In that situation, the luminosity of a $D$-dimensional black hole would be that of a perfect blackbody:
\begin{equation}\label{eq:luminosity-blackbody}
 L_{SB} \propto T^{D} \, .
\end{equation}

However, black holes should not be expected to behave as perfect blackbodies, given that emission of Hawking radiation effectively reduces the total mass available, shifting the location of the horizon \cite{fabbri2005modeling, torres2013evaporation}. To address this issue, we calculate the luminosity of an emitting black hole in the Parikh--Wilczek tunneling framework. This allows us to investigate how energy conservation and the adS geometry regulate the emission rate compared to an idealized blackbody. Considering the probability $\Gamma = e^{\Delta S}$ that is obtained by considering the backreaction from the emission of a single particle with energy $\omega$, we define the tunneling luminosity from the emission of multiple massless scalar particles as:
\begin{equation}\label{eq:Luminosity-tunneling}
L_{\text{tunnel}} = \frac{1}{2\pi} \int_0^{M(u)}  
\frac{\sigma(\omega, M, L) \, \omega}{e^{-\Delta S} - 1} \, d\omega \,,
\end{equation}
where $\sigma$ is a greybody factor. In the tunneling picture, it represents the probability that a particle, once created at the horizon, actually manages to escape the black hole's gravitational potential well to reach an observer at infinity. The upper integration limit at $M(u)$ is a strict statement of energy conservation. Unlike the Boltzmann distribution, which mathematically allows for the emission of particles at every energy $\omega$, the tunneling framework enforces that the black hole cannot radiate more energy than what it possesses at any\linebreak  given time. 

There are several approaches, with various degrees of accuracy, for calculating the greybody factor $\sigma$, considering that for adS black holes it depends on both $\omega$ and $L$. A solid approximation that covers both the large and small black-hole regimes is given by the geometric cross-section:
\begin{equation}
\sigma_{\text{geom}} =  \frac{\pi r_{ph}^2}{f(r_{ph})} =  \frac{\pi r_{ph}^2}{1 - \frac{2M}{r_{ph}} + \frac{r_{ph}^2}{L^2}} \, ,
\end{equation}
where $r_{ph}$ is the radius of the photon sphere. For a stationary SadS black hole, the photon sphere is located at $r_{ph} = 3M$. For small adS black holes ($M \ll L$): $\sigma \to 27 \pi M^2$,  matching the standard Schwarzschild result, whereas for large adS black holes ($M \gg L$): $\sigma \to 3 \pi L^2$ (the cross-section becomes proportional to the area of the adS radius).

When a particle tunnels out of the horizon, it does not immediately enter free space. It must traverse the curved spacetime potential barrier, and depending on the energy, it might be reflected by this barrier and fall back into the black hole. High-energy particles ``see'' the barrier as thin and pass through easily, whereas low-energy particles ($\omega \to 0$) are almost entirely reflected. To account for this, we use a frequency-dependent greybody factor $\sigma(\omega)$. We approximate this using the geometric cross-section modified by a Glauber-type filter:
\begin{equation}
\sigma(\omega, M, L) 
= \sigma_{\text{geom}} \times \bigg[ 1 - \exp\left( -\frac{\omega}{\omega_c} \right) \bigg]^n \,,
\end{equation}
where $\omega_c$ is a cutoff-energy depending on the height of the effective potential. As $M$ decreases, the photon sphere radius $r_{ph}$ shrinks, causing the ``filter'' to shift toward higher frequencies. This effectively ``blue-shifts'' the required energy for a particle to escape the adS barrier.

Our results demonstrate that the relationship between luminosity and mass depends fundamentally  on the adS scale $L$. In the large black-hole regime ($M \gg L$), the system behaves as a stable thermodynamic object. As the sigmoid function $M(u)$ decreases, both the temperature and the horizon area drop. The tunneling luminosity closely follows the Stefan--Boltzmann law because the available mass-energy $M$ is much larger than the typical energy ($\omega$) of a radiated particle. However, in the small black-hole regime ($M < L$), we observe a radical departure. While the temperature $T$ rises as $M$ decreases (mirroring Schwarzschild behavior), the luminosity does not diverge. Instead, it reaches a local peak and then drops to zero.

\begin{figure}[h]

\includegraphics[height=5cm,width=.45\linewidth]{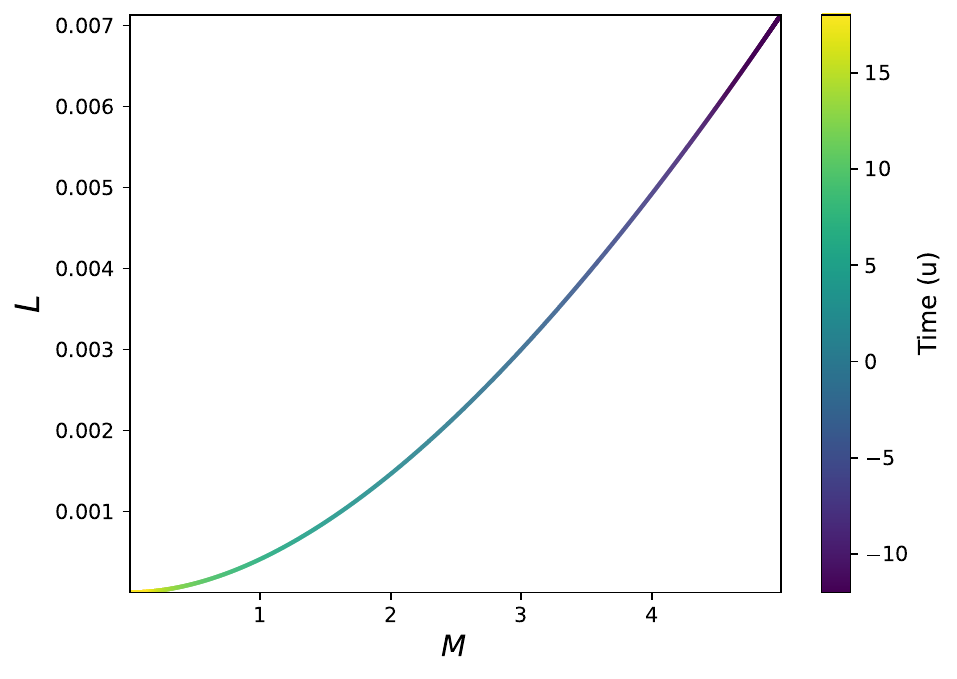}
\hspace{5mm}
\includegraphics[height=5cm,width=.45\linewidth]{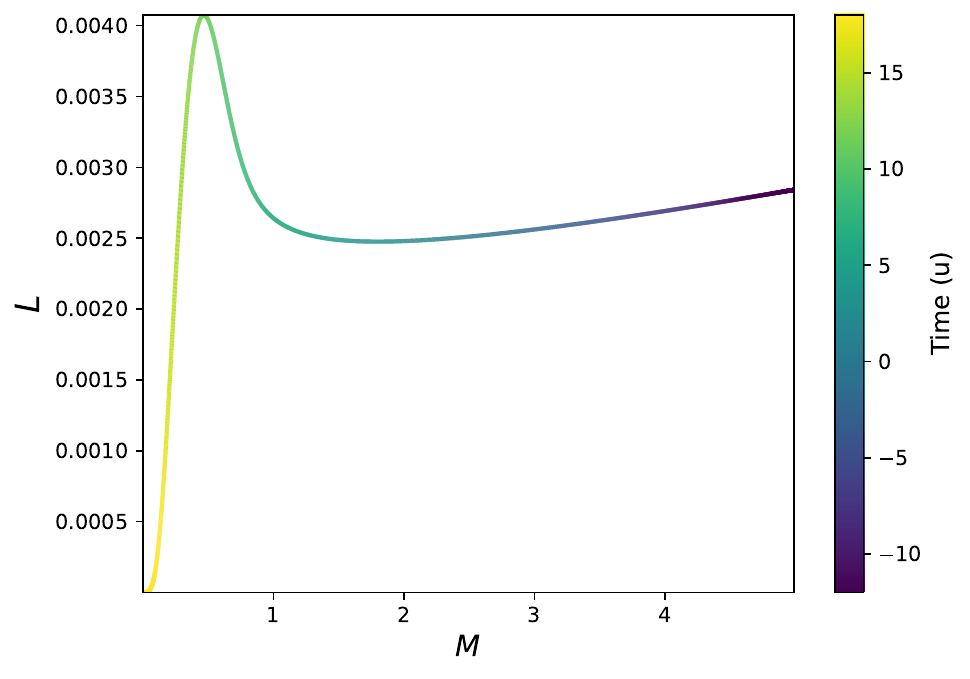}
\caption[Luminosity]{(\textbf{Left panel}) Black-hole luminosity as a function of its mass, for a large black hole ($M_0 = 5$ and $L=0.1$).
(\textbf{Right panel}) Black-hole luminosity as a function of its mass, for a small black hole ($M_0 = 5$ and $L=20$).}
\label{fig2}
\end{figure}

Figure~\ref{fig2} illustrates this framework. The graphs show that, rather than exhibiting the Boltzmann behavior where the luminosity is expected to diverge as the black hole loses mass, there is a drop corresponding to the loss of phase space for emission. This drop is a purely quantum backreaction effect. Although the black hole is hotter, it has so little remaining mass that the phase space for emission collapses. That is, the domain of integration in Equation~\eqref{eq:Luminosity-tunneling} is much reduced. The energy conservation term dominates over the thermal term, preventing the sharp increase predicted by the $L \propto T^{D}$ law ($L \propto T^{4}$ in the standard four-dimensional case).

\begin{figure}[h]

\includegraphics[height=5cm,width=.45\linewidth]{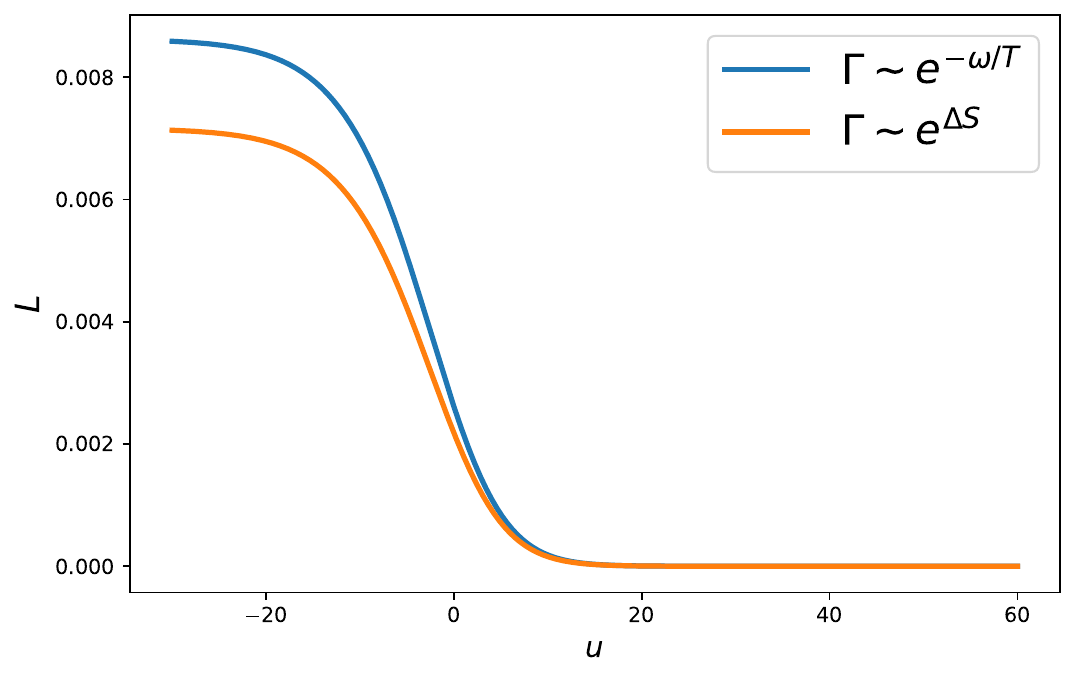}
\hspace{5mm}
\includegraphics[height=5cm,width=.45\linewidth]{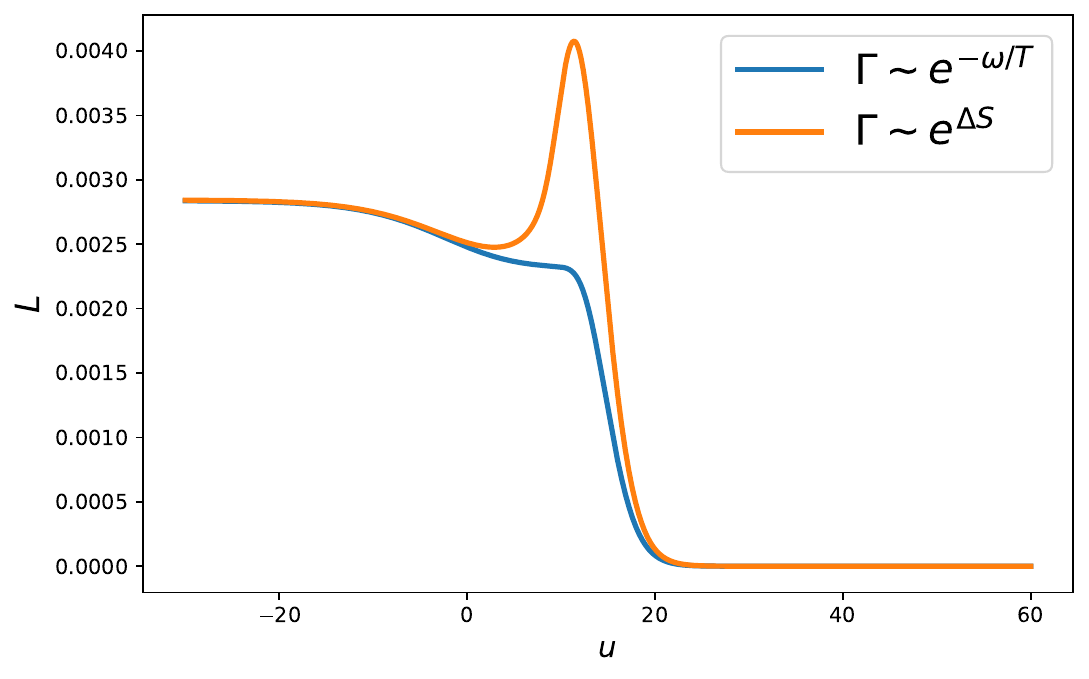}
\caption{Comparison between luminosity for the Boltzmann distribution and the tunneling distribution for an initial mass $M_0=5$. The left panel illustrates the large black-hole regime (setting $L = 0.1$). The right panel illustrates the small black-hole regime (setting $L = 20$).}
\label{fig3}
\end{figure}

We can compare the results obtained for the tunneling process with the behavior expected from thermal radiation with Boltzmann factor $\Gamma \sim e^{-\omega/T}$ by examining the time-dependent behavior of the luminosity function. Figure~\ref{fig3}  (left panel) shows the time evolution of the tunneling process of a large adS black hole ($M \gg L$). In this case, the time-dependent behavior reasonably approximates the Stefan--Boltzmann distribution. For large adS black holes, both the temperature and the horizon area decrease as the black hole radiates energy. In the right panel of Figure~\ref{fig3}, we present the same analysis for the tunneling process of a small adS black hole ($M < L$). Although the temperature increases as the black hole radiates, the luminosity from the tunneling process does not increase as the temperature rises. Considering the tunneling distribution $\Gamma \sim e^{\Delta S}$, we observe a significant peak in the emitted radiation before the collapse of the mass function $M(u)$, indicating the impact of the backreaction on the metric.

\section{Renormalized Energy--Momentum Tensor}
\label{section4}

A complete description of the evaporation process of a black hole through emission of Hawking radiation would require considering the spacetime backreaction, where the energy--momentum tensor of the radiation modifies the background geometry. As an approximation, the Hawking radiation emitted by a black hole can be treated as a atmosphere of particles with an energy--momentum tensor acting as a source on Einstein field equations \cite{t1998self,bardeen1981black}. In the semiclassical limit, the classic gravitational field couples to the expected value of the energy--momentum tensor of quantized matter fields via Einstein's field equations:
\begin{equation}
G_{\mu\nu} + \Lambda \,g_{\mu\nu} = 8\pi \langle T_{\mu\nu}\rangle. 
\end{equation}

A simple model of the Hawking atmosphere comes from the dynamics of a massless scalar field $\Phi$ satisfying the massless Klein--Gordon equation Equation~\eqref{eq:covariant_KG}, as discussed in Section~\ref{sec:Tunneling}. 
For a quantized field, the energy--momentum tensor $\langle T_{\mu\nu}\rangle$ is a divergent quantity and a renormalization scheme must be implemented. General expressions for the renormalized energy--momentum tensor in four dimensions are difficult to implement. For a massless scalar field, the four-dimensional energy--momentum tensor is: 
\begin{equation}
\langle T_{\mu\nu}\rangle^{(4D)} = \alpha R^2 + \beta R_{\mu\nu}R^{\mu\nu} + \gamma R_{\mu\nu\rho\sigma}R^{\mu\nu\rho\sigma} \,,
\end{equation}
where $R_{\mu\nu\rho\sigma}$, $R_{\mu\nu}$ and $R$ represent the Riemann tensor, the Ricci tensor and the Ricci scalar, respectively. In spherically symmetric spacetimes, it is possible to implement an optic geometric approximation. In this framework, the four-dimensional renormalized tensor is related to a two-dimensional effective tensor by \cite{ref-christensen} 
\begin{equation}
{\langle T_{\mu\nu}\rangle^{(4D)} = \frac{1}{4\pi r^{2}}{\langle T_{\mu\nu}\rangle^{(2D)}}}\,.
\end{equation}

In order to calculate the components of the renormalized two-dimensional energy momentum tensor, we follow the procedure outlined in \cite{beciu1984evaporating}. While the trace of a classical massless scalar field vanishes, the quantized field in two dimensions exhibits a conformal anomaly proportional to the scalar curvature:
\begin{equation}
\langle {{T^{\mu}}_{\mu}} \rangle^{(2D)} = \frac{R}{24\pi} \,.
\end{equation}

If we consider the $(u,r)$ part of the Vaidya-adS metric, the Ricci scalar is $R = f''(u, r)$, where the prime denotes $\p_r$. With this, the trace equation becomes:
\begin{equation}
  2 \langle T_{ur}\rangle - f(u,r) \langle T_{rr} \rangle = \frac{1}{24\pi} \left[ \frac{4M(u)}{r^3} - \frac{2}{L^2}\right] \, .  
\end{equation}

The components of the renormalized tensor $\langle T_{\mu\nu} \rangle$ must satisfy the the conservation law $\nabla_{\mu} \langle{T_{\nu}}^{\mu}\rangle = 0$. We focus on the component $\langle T_{uu} \rangle$, representing the energy density of the scalar field. Solving the radial differential equation for this component yields the\linebreak  following expression:
\begin{equation}
\langle T_{uu} \rangle = \frac{1}{24\pi}\left[ \frac{1}{2} f f'' - \frac{1}{4} (f')^2 + \frac{2\dot{M}(u)}{r^2} \right] + C\,,    
\end{equation}
where $C$ is an integration constant (with respect to the radial variable). In this case, we adopt $C = 0$, implying that the radiation reaching infinity does not bounce back, effectively allowing the black hole to evaporate.

At the adS boundary ($r \to \infty$), the metric function $f(u,r)$ is dominated by the $r^2/L^2$ term, and
\begin{equation}
\langle T_{uu} \rangle^{\text{boundary}} = \frac{1}{24\pi}\left[ \frac{1}{2} \left( \frac{r^2}{L^2} \right) \left( \frac{2}{L^2} \right) - \frac{1}{4} \left( \frac{2r}{L^2} \right)^2 + \frac{2\dot{M}(u)}{r^2} \right] =  \frac{\dot{M}(u)}{12\pi r^2}   \,.
\end{equation}

While the energy density component goes to zero as $r \to \infty$, the total luminosity (energy per unit time) passing through the boundary is:
\begin{equation}
L^{\text{boundary}} = \lim_{r \to \infty} \int \langle T_{uu} \rangle r^2 d\Omega =  \frac{\dot{M}(u)}{3}\,.       
\end{equation}
This shows that information propagation reaching the boundary is directly proportional to the rate at which the black hole is shrinking.

At the horizon, since $f(u,r_{H}) = 0$, we have
\begin{equation}\label{T_uu_horizon}
\langle T_{uu} \rangle \big|_{r_H} 
= \frac{1}{24\pi} \bigg\{- \frac{1}{4}[ f'(u,r_H)]^2 + \frac{2 \dot{M}(u)}{r_H^2}\bigg\}  
= -\frac{\pi T^2}{6} + \frac{\dot{M}(u)}{12\pi r_H^2} \,.
\end{equation}

The first term represents the expected (negative) energy density from thermal particles propagating in two dimensions. Since it is negative, it indicates that energy is flowing into the black hole from the vacuum, compensating for the radiation escaping to infinity. The second term, proportional to $\dot{M}$, represents the backreaction effect on the metric.
Because $\dot{M} < 0$, it also contributes to the inflow of negative energy density; the magnitude of which increases with a faster mass loss rate.

To conclude this section, we analyze the luminosity due to the energy--momentum  passing through the horizon area. Using Equation~\eqref{T_uu_horizon}, we obtain 
\begin{equation}
L\big|_{r_H} = (4\pi r_{H})^2 \langle T_{uu} \rangle \big|_{r_H}
= -\frac{2\pi^2 T^2}{3} + \frac{\dot{M}(u)}{3} \,.
\end{equation}

\begin{figure}[h]

\includegraphics[height=5cm,width=.45\linewidth]{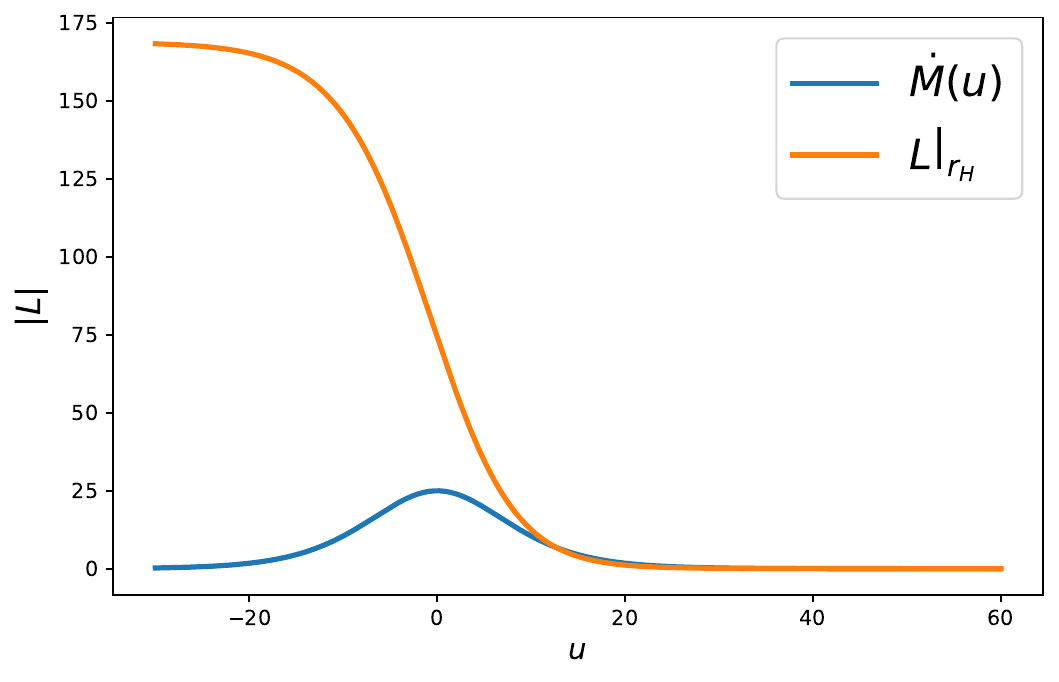}
\hspace{5mm}
\includegraphics[height=5cm,width=.45\linewidth]{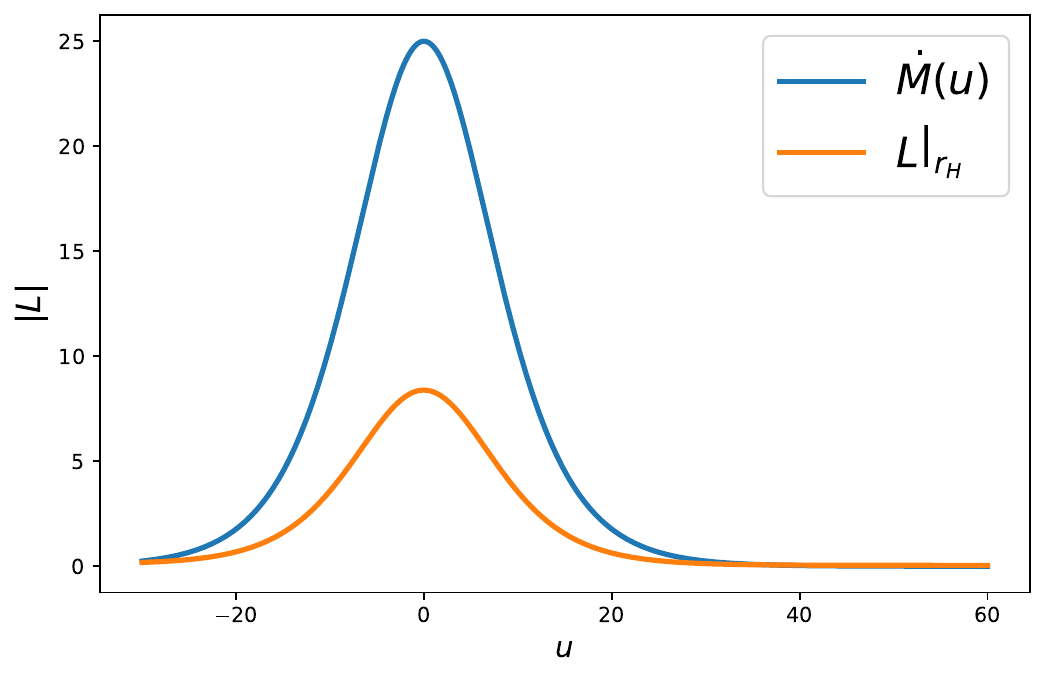}
\caption{Comparison between (absolute values of) luminosity evaluated at the horizon and mass loss rate for an initial mass $M_0=5$ The left panel illustrates the large black-hole regime (setting $L = 0.1$). The right panel illustrates the small black-hole regime (setting $L = 20$).}
\label{fig4}
\end{figure}

Our results indicate that for a large adS black holes, the main contribution to the luminosity comes from the thermal term proportional to $T^2$, exhibiting a similar behavior to the results obtained using the tunneling approach (see left panel of Figure~\ref{fig2}). On the other hand, for small adS black holes $(M < L)$ the dominating term in the contribution is the radiative term, proportional to the mass-loss rate $\dot{M}(u)$, that is associated with the backreaction of the metric and the reduction in the black hole horizon. These results are illustrated in Figure~\ref{fig4}.

\section{Final Remarks}
\label{sec:conclusion}

In this work, we analyzed the evaporation of an spherically symmetric anti-de Sitter black hole. We modeled its Hawking atmosphere using a Vaidya-adS geometry whose dynamics are governed by a sigmoid mass function. We described backreaction effects by using the Parikh--Wilczek framework for the tunneling of scalar particles.
This approach allowed us to transition smoothly between the large and small black-hole regimes while maintaining a continuous description of the geometry and its associated thermodynamics.

By comparing the Stefan--Boltzmann law with the Parikh--Wilczek tunneling framework, we identified a crucial physical cutoff. While idealized models predict a divergent temperature and luminosity as the black-hole mass tends to zero, the tunneling approach, constrained by energy conservation, shows that the luminosity remains finite and eventually vanishes as the available phase space for emission contracts. 
Our results show that for small adS black holes, the luminosity fails to increase monotonically with temperature, highlighting the profound influence of rapid mass loss and geometric backreaction on the thermodynamic evolution of asymptotically adS spacetimes.

\newpage

We also employed a technique based on defining an effective and renormalized energy--momentum tensor. We demonstrated that the effective energy density contains a critical non-adiabatic term proportional to the mass-loss rate. This term confirms that for small adS black holes, the evaporation process creates a substantial luminosity increase during the steep transition of the sigmoid, after which the luminosity collapses. 

This dual approach allows us to characterize the time-dependent luminosity and identify departures from Stefan--Boltzmann-like behavior during the final stages of evaporation. In principle, this work could be extended to model Hawking atmospheres around rotating anti-de Sitter black holes, for which more elaborate thermodynamic descriptions exist. Alternatively, the methods could be applied to characterize the evaporation dynamics of asymptotically de Sitter spacetimes, which has implications for investigating black holes in cosmological scenarios. Research along these lines is underway.


\section*{Funding}
    
A.~F.~C. acknowledges the partial support of Funda\c{c}\~{a}o de Amparo \`{a}  Pesquisa do Estado de S\~{a}o Paulo FAPESP (Grant No.~2013/07414-5).

\end{document}